%% file: arxiv.tex
\documentclass[aps,amsmath,amssymb,reprint,floatfix,longbibliography,prapplied]{revtex4-1}
\usepackage{graphicx}
\usepackage{dcolumn}
\usepackage{bm}
\usepackage[pdfencoding=auto, psdextra]{hyperref}
\usepackage{comment}
\usepackage[version=4]{mhchem}
\usepackage{color}
\usepackage{listings}

\usepackage{siunitx}
\DeclareSIPostPower\tothefourth{4}
\usepackage[english]{babel}
\usepackage{esdiff} 
\usepackage{mathtools}
\DeclarePairedDelimiter\abs{\lvert}{\rvert} 
\addto\captionsenglish{}
\addto\captionsenglish{}

\begin{document}
\title{Probing the current-phase relation of graphene Josephson junctions using microwave measurements}
\author{Felix E. Schmidt}
\author{Mark D. Jenkins}
\affiliation{Kavli Institute of NanoScience, Delft University of Technology, Lorentzweg 1, 2628 CJ, Delft, The Netherlands.}
\author{Kenji Watanabe}
\author{Takashi Taniguchi}
\affiliation{National Institute for Materials Science, 1-1 Namiki, Tsukuba, 305-0044, Japan}
\author{Gary A. Steele}\email[]{g.a.steele@tudelft.nl}
\affiliation{Kavli Institute of NanoScience, Delft University of Technology, Lorentzweg 1, 2628 CJ, Delft, The Netherlands.}


\begin{abstract}

We perform extensive analysis of graphene Josephson junctions embedded in microwave circuits.
By comparing a diffusive junction at \SI{15}{\milli\kelvin} with a ballistic one at \SI{15}{\milli\kelvin} and \SI{1}{\kelvin}, we are able to reconstruct the current-phase relation.
\end{abstract}

\maketitle

\section{Introduction}

Josephson junctions (JJs) are widely used in microwave (MW) applications, such as quantum limited amplification and sensing, where JJs are exploited as nonlinear inductors.
For the use of JJs in superconducting quantum information circuits, the junction nonlinearity has a major effect on the circuit requirements and capabilities~\cite{kringhojAnharmonicitySuperconductingQubit2018}.
However, the exact Josephson inductance can significantly differ between junctions:
While JJs are generally non-linear elements, the specific non-linearity depends on the current-phase relation (CPR) which in turn is determined by the underlying physics inside the junction.

The current-phase relation is a fundamental property of the JJ, relating the supercurrent $I_\text{J}$ flowing across a weak link between two superconducting banks with the phase difference $\delta$ between the two superconductors.
It results from the first derivative of the Josephson energy potential with respect to phase, $I_\text{J}(\delta) = (2e/\hbar) \partial_\delta V(\delta)$.
For the ideal case of a JJ formed by a thin insulating tunnel barrier between two superconducting electrodes (SIS), the Josephson potential is given by $V(\delta)/E_\text{J}=1-\cos\delta$ and the CPR has pure sinusoidal character as given by the first Josephson relation, $I_\text{J}(\delta) = I_\text{c}\sin\delta$~\cite{josephsonPossibleNewEffects1962,josephsonSupercurrentsBarriers1965}.

However, in JJs formed by normal conductors between superconductors (SNS) such as graphene Josephson junctions (gJJs), transport across the JJ is governed by Andreev bound states (ABS), each with ground state energy
\begin{align}
	V_i(\delta)/\Delta_0=1-\sqrt{1-\tau_i\sin^2(\delta/2)}
	\label{CPReq:ABSenergy}
\end{align}
with transmission probability $\tau_i$ and superconducting gap $\Delta_0$~\cite{beenakkerUniversalLimitCriticalcurrent1991,titovJosephsonEffectBallistic2006b}.
Assuming a JJ with $N$ channels of equal $\tau_i$, i.e. $\tau=\sum\tau_i/N$, the corresponding CPR is given by
\begin{align}
	I_\text{J}(\delta) = \frac{\pi\Delta_0}{2 e R_\text{n}} \frac{\sin\delta}{\sqrt{1 - \tau \sin^2(\delta / 2)}},
	\label{CPReq:CPR-ball}
\end{align}
with the Boltzmann constant $k_\text{B}$ and normal state resistance $R_\text{n}= R_\text{q}/N = h/(Ne^2)\approx \SI{25.812}{\kilo\ohm} / N$~\cite{golubovCurrentphaseRelationJosephson2004a,leeUltimatelyShortBallistic2015}.
Here, $R_\text{q}$ denotes the quantum Hall resistance and $N$ the number of conducting channels.
Depending on $\tau$, the CPR can exhibit significant forward skew compared to the case of a purely sinusoidal CPR in SIS JJs.
While the CPR of gJJs has been studied in the DC regime~\cite{englishObservationNonsinusoidalCurrentphase2016,nandaCurrentPhaseRelationBallistic2017}, and gJJs have been successfully incorporated in MW circuits~\cite{schmidtBallisticGrapheneSuperconducting2018,krollMagneticFieldCompatible2018,wangCoherentControlHybrid2019}, the influence of the potentially skewed CPR has not been studied in the latter.

Here, we analyze the effect of a nonlinear CPR on the microwave performance of gJJ embedded in microwave circuits.
Measuring two devices in different states, we compare the influence of scattering transport and temperature on the JJ nonlinearity.
Our circuit design allows in-situ, and even simultaneous, DC and MW measurements, providing us with various measurement types to compare.
The results show the usefulness of combining DC and MW in the same circuits for fundamental research on Josephson junction physics, which distinguishes it from pure MW CPR measurements~\cite{rifkinCurrentphaseRelationPhasedependent1976}.

\section{Circuit characterization}

\begin{figure*}[t]
	\centering
	\includegraphics[width=.7\linewidth]{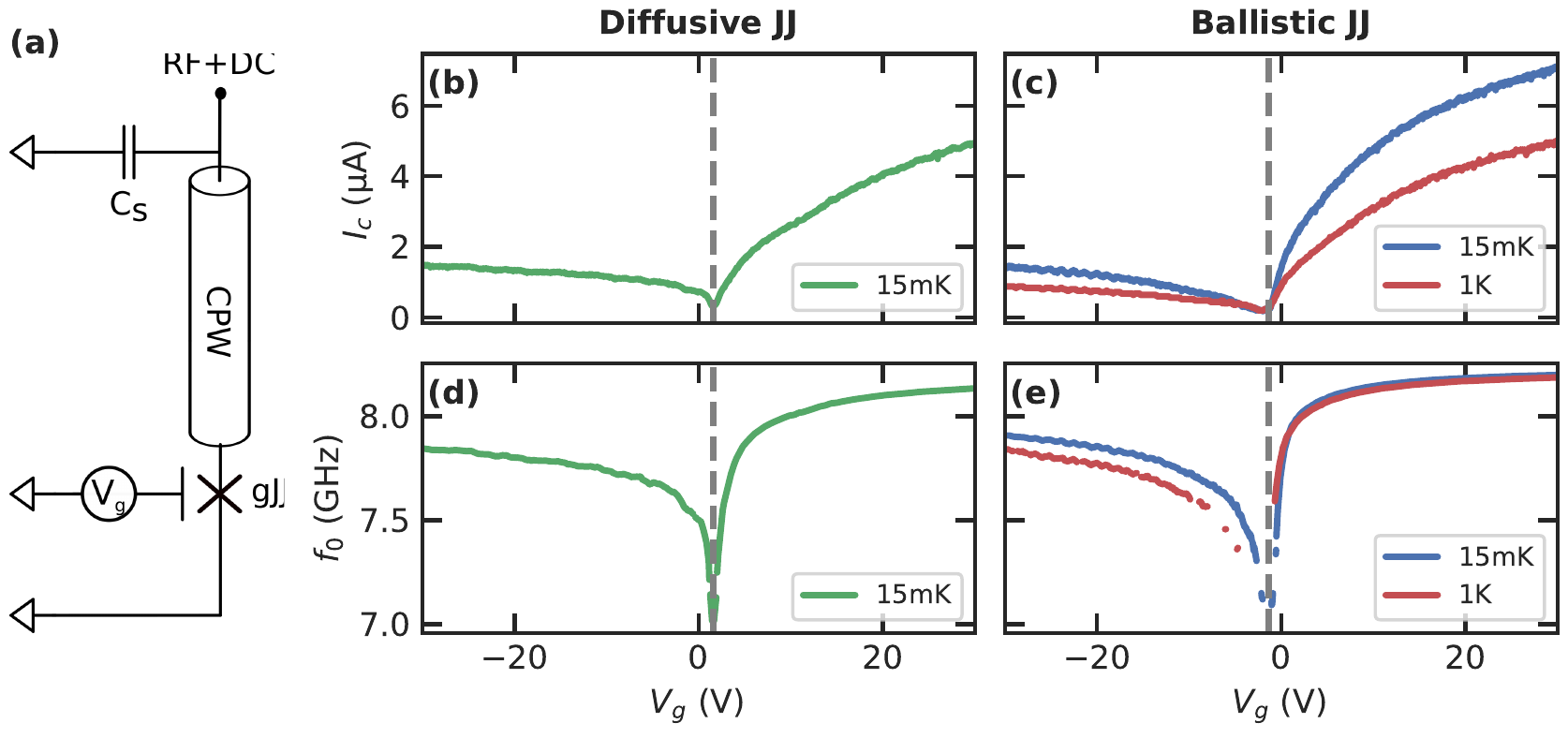}
	\caption{
		\textbf{Simultaneous MW and DC measurements of ballistic and diffusive graphene Josephson junctions.}
		\textbf{(a)} Measurement schematic.
		The gJJ shorts a coplanar waveguide transmission line to ground, which forms a gate-tunable $\lambda/2$-resonator.
		$V_\text{g}$ is fed through an additional shunt capacitor (not shown).
		\textbf{(b,c)} Switching current for the diffusive \textbf{(b)} and ballistic Josephson junction \textbf{(c)}, at base-temperature of \SI{15}{\milli\kelvin} (blue) and at \SI{1}{\kelvin} (red).
		\textbf{(d,e)} Resonance frequencies versus gate voltage for the diffusive \textbf{(d)} and ballistic \textbf{(e)} device.
		The gate-tunable Josephson inductance changes the boundary condition of the $\lambda/2$-resonator, thus changing the resonance frequency of the circuit.
		Dashed grey lines indicate the charge neutrality point of each device, marked by the minimum critical current.
	}
	\label{CPRfig:figure1}
\end{figure*}

Our circuit consists of a DC-bias microwave cavity formed by a coplanar waveguide (CPW) which is shunted by a large capacitor at the input, and shorted to ground on the far end by a gJJ that can be tuned with a gate voltage ($V_\text{g}$), see Fig.~\ref{CPRfig:figure1}(a) and Refs.~\cite{schmidtBallisticGrapheneSuperconducting2018,schmidtCurrentDetectionUsing2020,bosmanBroadbandArchitectureGalvanically2015c}.
The superconducting base layer and shunt capacitor metal layers consist of DC-sputtered molybdenum-rhenium on a sapphire substrate, while the shunt capacitor dielectric layer is PECVD-\ce{SiN_x}.
The gate voltage lead is fed through a second shunt capacitor of the same geometry as the one at the input in order to suppress MW radiation leaking in through or out of the gate line.
The MW wiring of both samples was fabricated on a single \SI{2}{inch} sapphire wafer, after which the wafer was diced into \SI{10x10}{\milli\meter} pieces onto which the individual gJJ were placed.
The gJJ consist of boron nitride encapsulated single layer graphene with side-contacts of DC-sputtered niobium titanium nitride (NbTiN), fabricated via the etch-fill technique~\cite{wangOneDimensionalElectricalContact2013b,schmidtBallisticGrapheneSuperconducting2018}.
The gJJ are designed to be \SI{5}{\micro\meter} wide and separate the NbTiN leads by a length of \SI{500}{\nano\meter}.
Gate tunability is achieved by placing a third NbTiN lead extending over the entire gJJ, separated by a bilayer of HSQ.
The circuit is wirebonded into a PCB that is mounted on the millikelvin plate of a dilution refrigerator and connected to the outside world via a bias-T, allowing both DC and MW characterization in the same setup.
To suppress thermal excitations, the MW input line is heavily attenuated and all DC lines were equipped with $\pi$-filters in the room temperature battery powered electronics, as well as copper powder and two-stage RC filters thermally anchored to the millikelvin stage.

We measured two separate devices with nominally identical microwave circuits and junction designs:
One of the devices exhibited signatures of ballistic transport in form of Fabry-Pérot-like oscillations, which we will refer to as the \textit{ballistic device} (see Supplementary Material Sec.~\ref{sec:SMballistic} and Fig.~\ref{CPRfig:SMFig-ballistic}).
This is the device presented in the main text of Ref.~\cite{schmidtBallisticGrapheneSuperconducting2018}.
The other one, in lack of such features, will be called \textit{diffusive device}, and corresponds to the reference sample of Ref.~\cite{schmidtBallisticGrapheneSuperconducting2018}.
With a normal state resistance of both devices ranging between \SIrange{35}{350}{\ohm}, depending on gate voltage, we estimate around 74 to 740 conducting channels.
This justifies the use of a single averaged transparency parameter $\tau$ in Eq.~\eqref{CPReq:ABSenergy}.

We extract the DC circuit parameters by applying a bias current to the JJ, using the CPW as a long capacitive lead and measuring the voltage drop across the gJJ.
When exceeding a critical current, the JJ switches from the zero-voltage to the resistive state.
We record this switching current $I_\text{c}$ for varying gate voltages, as depicted in Fig.~\ref{CPRfig:figure1}(b,c) for the two devices at a base temperature of \SI{15}{\milli\kelvin} in the case of the diffusive, and both base temperature and \SI{1}{\kelvin} for the ballistic device.
In line with a minimum conductivity even at the CNP~\cite{katsnelsonZitterbewegungChiralityMinimal2006, zieglerRobustTransportProperties2006, tworzydloSubPoissonianShotNoise2006,titovJosephsonEffectBallistic2006b}, there remains a finite supercurrent in both samples, that cannot be pinched off completely.
The DC switching current of the diffusive device ranges from a few hundred \si{\nano\ampere} to \SI{5.5}{\micro\ampere}, similar to the ballistic device at \SI{1}{\kelvin}.
At base temperature, the maximum $I_\text{c}$ of the ballistic device reaches up to \SI{7.5}{\micro\ampere}.
Both samples exhibit significantly larger switching current for $V_\text{g}>V_{\rm CNP}$ (n-doping) compared to $V_\text{g}<V_{\rm CNP}$ (p-doping), where $V_{\rm CNP}$ denotes the gate voltage at the charge neutrality point (CNP) of the gJJ.
We attribute this to a reduced contact transparency in the p-doped regime~\cite{schmidtBallisticGrapheneSuperconducting2018}.
We measure $V_{\rm CNP}^{\rm diff}=\SI{1.55}{\volt}$ and $V_{\rm CNP}^{\rm ball}=\SI{-1.39}{\volt}$ for the diffusive and ballistic sample, respectively.
Discrepancies are presumably due to differences in residual doping during fabrication.

For high frequency signals, i.e. a few \si{\giga\hertz}, the gJJ behaves as a nonlinear inductor, with Josephson inductance
\begin{align}
	L_\text{J} = \frac{\hbar}{2e}\left(\diff{I_\text{J}}{\delta}\right)^{-1},
	\label{CPReq:LJgeneral}
\end{align}
which can be derived from the second Josephson relation, $\partial_t\delta=2e V/\hbar$.
The resonance frequency of a $\lambda/2$-resonator shorted to ground by such a Josephson inductance can be approximated by
\begin{align}
	f_0\left(I_\text{b},I_\text{c}\right) = f_{\lambda/2} \frac{L_\text{r}+L_\text{J}\left(I_\text{b}, I_\text{c}\right)}{L_\text{r} +  2L_\text{J}\left(I_\text{b}, I_\text{c}\right)}
	\label{CPReq:Pogorzalek}
\end{align}
with $L_\text{r}$ the bare CPW inductance and $f_{\lambda/2}$ the resonance frequency of the CPW without the JJ, see Supplementary Material Sec.~\ref{sec:SMcalibration}.
$I_\text{b}$ is the bias current flowing through CPW and the JJ, $I_\text{c}$ the critical current of the JJ.
Depending on the impedance of the gJJ at the circuit resonance frequency, $Z_\text{J}=i\omega_0 L_\text{J}$, the fundamental mode hosted by the gJJ-terminated CPW varies between a $\lambda/2$ wave ($f_0\rightarrow f_{\lambda/2}$) for small $Z_\text{J}\rightarrow0$, while for $L_\text{J}\gg L_\text{r}$ the fundamental mode is $\lambda/4$ ($f_0 \rightarrow f_{\lambda/2}/2=f_{\lambda/4}$).

The circuit response is measured by recording the reflection coefficient $S_{11}$ of the cavity using a vector network analyzer, which excites the device through a series of attenuators and a directional coupler, and measures the reflected signal, amplified by low noise cryogenic and room temperature HEMTs. 
We fit the response using an analytical model to extract resonance frequency $f_0$ and internal ($\kappa_\text{i}$) and external loss rates ($\kappa_\text{e}$), see Supplementary Material Sec.~\ref{sec:SMextraction}.
We observe gate-tunable resonance frequency $f_0$ between \SIrange{7.0}{8.2}{\giga\hertz}, comparable for both devices, see Fig.~\ref{CPRfig:figure1}(d,e).
Due to the inverse nature of junction current and inductance, the large changes in $I_\text{c}$ for $V_\text{g}>V_{\rm CNP}$ only lead to minor changes in $f_0$ when comparing the hot and cold ballistic device.
On the other hand, even small changes in the significantly smaller $I_\text{c}$ for $V_\text{g}<V_{\rm CNP}$ significantly reduce $f_0$ in this regime.

\section{Deviations between Josephson inductance from DC and MW measurements}

\begin{figure*}[t]
	\centering
	\includegraphics[width=.7\linewidth]{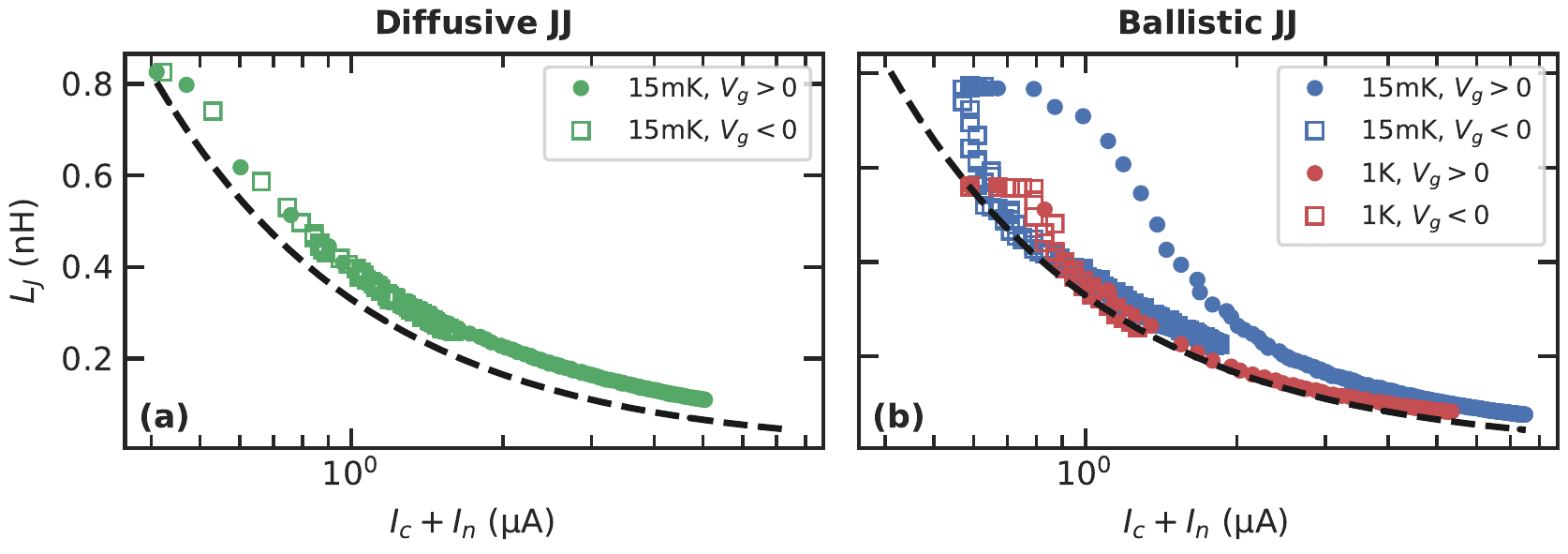}
	\caption{
		\textbf{Evidence for non-sinusoidal CPR from deviation between Josephson inductance and critical current.}
		MW-extracted $L_\text{J}$ versus DC-measured $I_\text{c}$, corrected for estimates of current noise of \SI{110}{\nano\ampere} for the diffusive device at \SI{15}{\milli\kelvin} \textbf{(a)} and \SI{390}{\nano\ampere} for the ballistic one at \SI{15}{\milli\kelvin} and at \SI{1}{\kelvin} \textbf{(b)} (blue and red, respectively).
		Full circles (empty squares) correspond to $V_\text{g}>V_{\rm CNP}$ ($V_\text{g}<V_{\rm CNP}$).
		Dashed line corresponds to an $L_\text{J}$ calculated from $I_\text{c}$ assuming a sinusoidal CPR.
		Values of $L_\text{J}$ above the dashed line indicate a forward-skewed CPR, values below the dashed line would correspond to backwards skewing.
	}
	\label{CPRfig:figure2}
\end{figure*}

Assuming a purely sinusoidal current-phase relation, the Josephson inductance can be extracted from the current phase relation via $L_\text{J} = \hbar/(2\pi I_\text{c} \cos\delta)$.
However, depending on the exact shape of the CPR, $L_\text{J}$, and with it $f_0$, can significantly deviate from the above equations, see Supplementary Fig.~\ref{CPRfig:SMinfluence}.
This leads to a reduced slope of the CPR around zero phase, which enhances $L_\text{J}$ compared to the case of a sinusoidal CPR for the same value of $I_\text{c}$.

Instead of relying only on the DC measured values of $I_\text{c}$ and the assumption of SIS CPR, we can directly extract $L_\text{J}$ from the MW measurement of $f_0$.
To calibrate the circuit parameters, we use additional measurements of reference devices shorted to with an open and a short to ground instead of a gJJ (see Supplementary Material Sec.~\ref{sec:SMcalibration} for details).
From this, we extract $f_{\lambda/2}=\SI{8.364}{\giga\hertz}$ and $L_\text{r}=\SI{3.671}{\nano\henry}$, which allows us to extract $L_\text{J}$ via Eq.~\eqref{CPReq:Pogorzalek}.

In Fig.~\ref{CPRfig:figure2}, we plot the observed Josephson inductance together with the measured critical currents for the measured devices.
As detailed in Supplementary Material Sec.~\ref{sec:SMfitbiascurrent}, we estimate low-frequency current noise $I_\text{n}$ to range between \SIrange{110}{390}{\nano\ampere} in the setups used for measuring the diffusive and ballistic device, respectively.
Without accounting for $I_\text{n}$, the observed $L_\text{J}$ is significantly smaller than the SIS-CPR estimate from DC measurements of $I_\text{c}$, which, without any current noise, could only be explained by a backward-skewed CPR, see Supplementary Fig.~\ref{CPRfig:SMfigure2}.
However, added to the measured values of $I_\text{c}$, this amount of current noise is sufficient to move all data points such that $L_\text{J}$ is larger than expected from sinusoidal CPR for all $I_\text{c}$, matching the expected forward-skewed CPR regardless of diffusive or ballistic transport, or elevated temperatures.

The deviation is largest for the ballistic device at base temperature, and significantly reduced for the diffusive device, or at \SI{1}{\kelvin}.
This matches with the expectation of reduced forward skewing of the CPR at higher temperatures or lower transparencies:
The skew is due to the phase coherence of Andreev bound states traversing the normal region between the superconducting banks multiple times (or, in a similar picture, multiple ABS crossing the normal region) which in turn means a longer phase coherence length is required to keep this contribution.
As the phase coherence length is highly sensitive to temperature and scattering, an increase in either one of the last two results in both a reduction of switching current and forward skewing~\cite{fuechsleEffectMicrowavesCurrentPhase2009,hagymasiJosephsonCurrentBallistic2010,black-schafferStronglyAnharmonicCurrentphase2010,rakytaMagneticFieldOscillations2016,englishObservationNonsinusoidalCurrentphase2016}.

In order to examine the underlying mechanisms further, we continue by studying the power and bias current dependence of our circuit.

\section{Pure MW measurements of the Josephson nonlinearity}

\subsection{Probing $L_\text{J}$ via the power dependence}

The nonlinear inductance of a Josephson junction consequently introduces nonlinear behavior to the overall circuit.
Depending on the exact circuit design and participation ratio between Josephson and total circuit inductance, this nonlinearity is more or less diluted, yet finite so-called anharmonicity $\beta$, i.e. deviation from the ideal case of pure LC-resonator behavior, remains.
Our circuit architecture allows us to extract this quantity directly and to calculate the expected CPR skew.

\begin{figure*}[t]
	\centering
	\includegraphics[width=.7\linewidth]{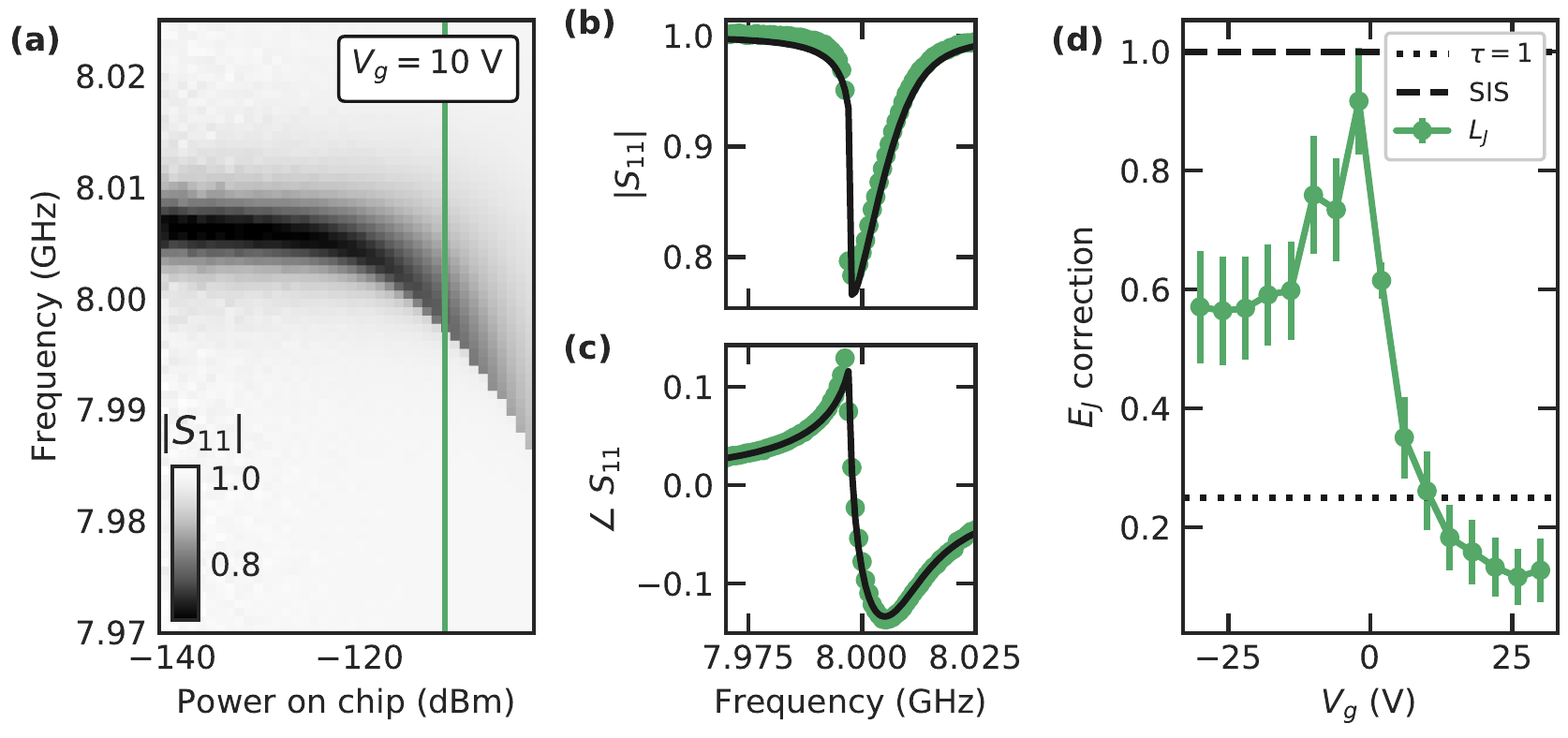}
	\caption{
		\textbf{Extracting the anharmonicity coefficient.}
		\textbf{(a)} Absolute value of the reflection coefficient $S_{11}$ versus frequency for increasing drive power.
		Due to the circuit nonlinearity, the resonator experiences a downshift and bifurcation at elevated drive powers.
		Solid lines indicate linecuts in \textbf{(b)} and \textbf{(c)}.
		\textbf{(b-c)} Absolute value \textbf{(b)} and phase \textbf{(c)} of $S_{11}$ for $P_\text{int}=\SI{-110}{dBm}$ as indicated in \textbf{(a)}.
		Black lines are fits.
		\textbf{(d)} Josephson energy correction for measured gate voltages.
		Dots: data as extracted from fits as in \textbf{(b-c)} and $L_\text{J}$, dashed line: SIS limit, dotted line: $\tau=1$ limit.
		Values below the dashed line indicate forward skewed CPR.
	}
	\label{CPRfig:figure3}
\end{figure*}

We can observe the anharmonicity of our DC bias circuit terminated with the diffusive gJJ by performing $S_{11}$ measurements at high drive powers for a series of different gate voltages, as shown in Fig.~\ref{CPRfig:figure3} for $V_\text{g}=\SI{+10}{\volt}$.
At very low drive powers, $\beta$ has negligible effect on the circuit response, which can still be described by a purely harmonic oscillator here.
With increasing on-chip power $P_{\rm in}$, the resonance frequency experiences a down-shift, and both amplitude and phase of $S_{11}$ start to get skewed towards lower frequencies.
Once $P_{\rm in}$ exceeds a critical threshold, the resonator response bifurcates, which can be seen by the discontinuity in the data.
For reference, all other measurements of this device were performed at $P_{\rm in}\approx\SI{-131.4}{dBm}$, still in the linear regime and with a maximum current at the junction of $I_{\rm MW}\approx\SI{3.0}{\nano\ampere}$ well below the critical current, see Supplementary Material Fig.~\ref{CPRfig:SMFigcurratJJ}.

Using the previously determined parameters $f_0$, $\kappa_\text{i}$ and $\kappa_\text{e}$, we can model the data by solving the equation of motion of a harmonic oscillator with an additional third order term in the cavity field with amplitude $\beta$, 
\begin{align}
	\dot{\alpha} = \left[ -i \left( \Delta+\beta\abs{\alpha}^2 \right)-\frac{\kappa}{2} \right]\alpha + \sqrt{\kappa_\text{e}} S_\text{in}\ ,
	\label{CPReq:Duffing-EOM}
\end{align}
where $S_\text{in}$ is the field amplitude of the drive, $\Delta$ the frequency detuning and $\kappa=\kappa_\text{i}+\kappa_\text{e}$, as detailed in Supplementary Material Sec.~\ref{sec:SMduffing}.

Best agreement between data and model is reached when introducing nonlinear dissipation in the form of increasing internal linewidth that grows with the square root of the drive power, $\delta\kappa_\text{i}/\kappa_\text{i}(0)=\gamma \sqrt{P_{\rm in}}$, see Supplementary Section~\ref{sec:SMduffing} and Supplementary Figs.~\ref{CPRfig:SMpower} and \ref{CPRfig:SMFig-lossrates-power}.
This is in contrast with circuits incorporating standard aluminum oxide JJs, where nonlinear dissipation with increasing power is usually absent~\cite{boakninDispersiveMicrowaveBifurcation2007b}.

There are several dissipation mechanisms known in superconducting microwave circuits that depend on drive power, such as on-chip heating~\cite{portisPowerinducedSwitchingHTS1991,heinFundamentalLimitsLinear1997,wosikPowerHandlingCapabilities1997}, dielectric losses~\cite{martinisDecoherenceJosephsonQubits2005c,oconnellMicrowaveDielectricLoss2008a,gunnarssonDielectricLossesMultilayer2013,lisenfeldElectricFieldSpectroscopy2019}, or subgap losses~\cite{dassonnevilleDissipationSupercurrentFluctuations2013,ferrierPhasedependentAndreevSpectrum2013,dassonnevilleCoherenceenhancedPhasedependentDissipation2018}.
Heating of the circuit itself is unlikely since $f_0$ should tune significantly stronger due to a reduced $I_\text{c}$ at elevated temperatures, with potentially significant influence on $f_0$, c.f. Fig.~\ref{CPRfig:figure1}, which we did not observe for any of the gate voltages.
Moreover, the power dissipated on-chip is extremely small and very unlikely to cause even local heating.

Losses due to electric dipole moments of two-level systems are also unlikely the source of the observation, as these are known to be activated for decreasing drive excitation voltages~\cite{martinisDecoherenceJosephsonQubits2005c,oconnellMicrowaveDielectricLoss2008a,gunnarssonDielectricLossesMultilayer2013}.
Moreover, TLS mainly reside in disordered dielectric materials.
However, there is only dielectric volume present at the shunt capacitor dielectric and the gJJ (encapsulating BN and HSQ top-gate).
Here, the circuit has voltage nodes and voltage fluctuations, which could activate the TLS, are expected to have negligible effect on the circuit performance.

We therefore attribute the source of the observed nonlinear damping to low-lying subgap states within the induced superconducting gap in the gJJ.
These subgap states can be due to e.g. intransparent superconductor-normal contacts, or Andreev bound states with large transverse momentum, polluting the bulk superconducting gap and leading to microwave loss~\cite{schmidtBallisticGrapheneSuperconducting2018}.
As the drive power increases, these subgap states get populated, resulting in an internal loss rate that grows with the square root of the input power, see Supplementary Fig.~\ref{CPRfig:SMFig-lossrates-power}.
Loss mechanisms in similar SNS systems, with normal metal weak links, have shown similar effects~\cite{fuechsleEffectMicrowavesCurrentPhase2009,dassonnevilleDissipationSupercurrentFluctuations2013}, but they have not been observed before in gJJ.

The $\beta$ term in Eq.~\eqref{CPReq:Duffing-EOM} is due to the anharmonicity of the microwave cavity for high drive powers which is evident when expanding the Josephson energy potential to higher orders,
\begin{align}
	V_J(\delta) \approx E_\text{J} \frac{\delta^2}{2} - E_\text{J}\left( 1-\frac{3\sum\tau_i^2}{4\sum\tau_i} \right) \frac{\delta^4}{24} +\mathcal{O}(\delta^6)\ , 
	\label{CPReq:EJtaylor}
\end{align}
where $E_\text{J}=\Delta_0\sum\tau_i/4$~\cite{kringhojAnharmonicitySuperconductingQubit2018}.
Compared to the case of an SIS junction, depending on $\tau$ the fourth-order correction 
\begin{align}
	\Gamma = 1-3\tau/4
	\label{CPReq:Ejcorrection}
\end{align} 
can vary between 1 for SIS to 0.25 for $\tau=1$.
In Fig.~\ref{CPRfig:figure3}(d), we plot this quantity as the ratio of the measured value of the anharmonicity coefficient $\beta_{\rm meas}$ and the one expected from a $\lambda/2$ resonator shorted to ground by a Josephson junction, approximately given by $\beta_{\rm th}=f_0p^3/2$ with the participation ratio between Josephson and total inductance $p=L_\text{J}/(L_\text{r}+L_\text{J})$~\cite{wilsonPhotonGenerationElectromagnetic2010b,zhouHighgainWeaklyNonlinear2014}.

For a broad range of gate voltages, the correction lies between the two extremes of no and full forward skewing.
However, for $V_\text{g}>\SI{5}{\volt}$, this value drops below the minimum of 0.25 as expected from Eq.~\eqref{CPReq:EJtaylor}.
While this is unexpected, we note that without knowing exactly how many ABS channels are active in the JJ, it is not possible to extract a number for $\tau$, as the measured anharmonicity coefficient only returns information on $\sum\tau_i=N\tau$.
Additional experiments, such as extracting the transparency for each channel from multiple Andreev reflection via voltage-biased measurements~\cite{scheerConductionChannelTransmissions1997,goffmanConductionChannelsInAsAl2017,bretheauTunnellingSpectroscopyAndreev2017a,pandeyAndreevReflectionBallistic2019}, or direct measures of both $E_\text{J}$ and the anharmonicity coefficient in transmon qubits~\cite{larsenSemiconductorNanowireBasedSuperconductingQubit2015, delangeRealizationMicrowaveQuantum2015, casparisGatemonBenchmarkingTwoQubit2016a,casparisSuperconductingGatemonQubit2018}, would be required to draw further conclusions.

\subsection{Probing $L_\text{J}$ via the bias current dependence}

A second way of reconstructing the CPR is by means of analyzing the bias current dependence of the high frequency circuit response, as this allows for a direct measure of $L_\text{J}(I_\text{b})$.
We model the bias current dependence of both the ballistic and diffusive device at \SI{15}{\milli\kelvin} using Eqs.~\ref{CPReq:LJgeneral} and \ref{CPReq:Pogorzalek} under the assumption of a general CPR according to Eq.~\eqref{CPReq:CPR-ball} and using $\tau$ and $I_\text{c}$ as a free parameters, as shown in Fig.~\ref{CPRfig:figure4}(a) (see Supplementary Material Sec.~\ref{sec:SMfitbiascurrent} for details).

Compared to a Josephson inductance with sinusoidal CPR, the measured data requires additional Josephson inductance, pushing $f_0$ to lower frequencies, which is provided by a CPR with same $I_\text{c}$, but forward skewed (see Supplementary Material Fig.~\ref{CPRfig:SMinfluence}).
The lower limit of the resonance frequency at zero bias current is given by a fully forward skewed CPR with $\tau=1$, which yields maximum $L_\text{J}$ for the same $I_\text{c}$ as a fully sinusoidal CPR.
For all gate voltages, the measured data lies between these two extremes.
Fixing $L_\text{r}$ and $f_{\lambda/2}$ as the earlier calibrated values, and including a forward skewed CPR in our model, we are able to fit the measured $f_0$, which allows us to extract a CPR-transparency parameter $\tau(V_\text{g})$.

As the bias current increases, so does the internal linewidth of the $S_{11}$ resonance, see Supplementary Material Sec.~\ref{sec:SMfitbiascurrent} and Fig.~\ref{CPRfig:SMFig-lossrates-current}.
This is most likely due to the previously mentioned current noise on our DC lines, which modulates the resonance frequency around the value set by $f_0$.
Due to the measurement time, the recorded trace then shows a widened resonance dip, that even resembles a split-dip feature at high responsivity to bias current, $G_1=\partial f_0/\partial I_\text{b}$.
We therefore chose to omit bias current measurements of gate voltages where the resonance frequency was not clearly visible, which is the reason for some missing datapoints in Fig.~\ref{CPRfig:figure4}.

From the remaining data, we extract an average channel transmission $\tau_{\rm diff}=0.64\pm0.18$ and $\tau_{\rm ball}=0.77\pm0.14$ for the diffusive and ballistic device, respectively, at base temperature.
With skew defined as the deviation of the CPR maximum from phase $\pi/2$, $S=2\delta_{\rm max}/\pi -1$, the corresponding values are $S_{\rm diff}=0.20\pm0.09$ and $S_{\rm ball}=0.27\pm0.15$ for the diffusive and ballistic device, respectively, as plotted in Fig.~\ref{CPRfig:figure4}(b).
We note that this is comparable to the results obtained from DC-measurements of the CPR~\cite{englishObservationNonsinusoidalCurrentphase2016,nandaCurrentPhaseRelationBallistic2017}.
Overall, the skewness seems to be constant for both devices, except for the region around CNP, where skewness seems to be significantly higher than elsewhere.

Corrections to $E_\text{J}$ amount to $0.52\pm0.13$ for the diffusive and $0.42\pm0.10$.
This is an important result for future use of gJJs in applications such as qubits, as this correction plays an important role in the circuit's anharmonicity and coherence times~\cite{kringhojAnharmonicitySuperconductingQubit2018}.

\begin{figure*}[t]
	\centering
	\includegraphics[width=.7\linewidth]{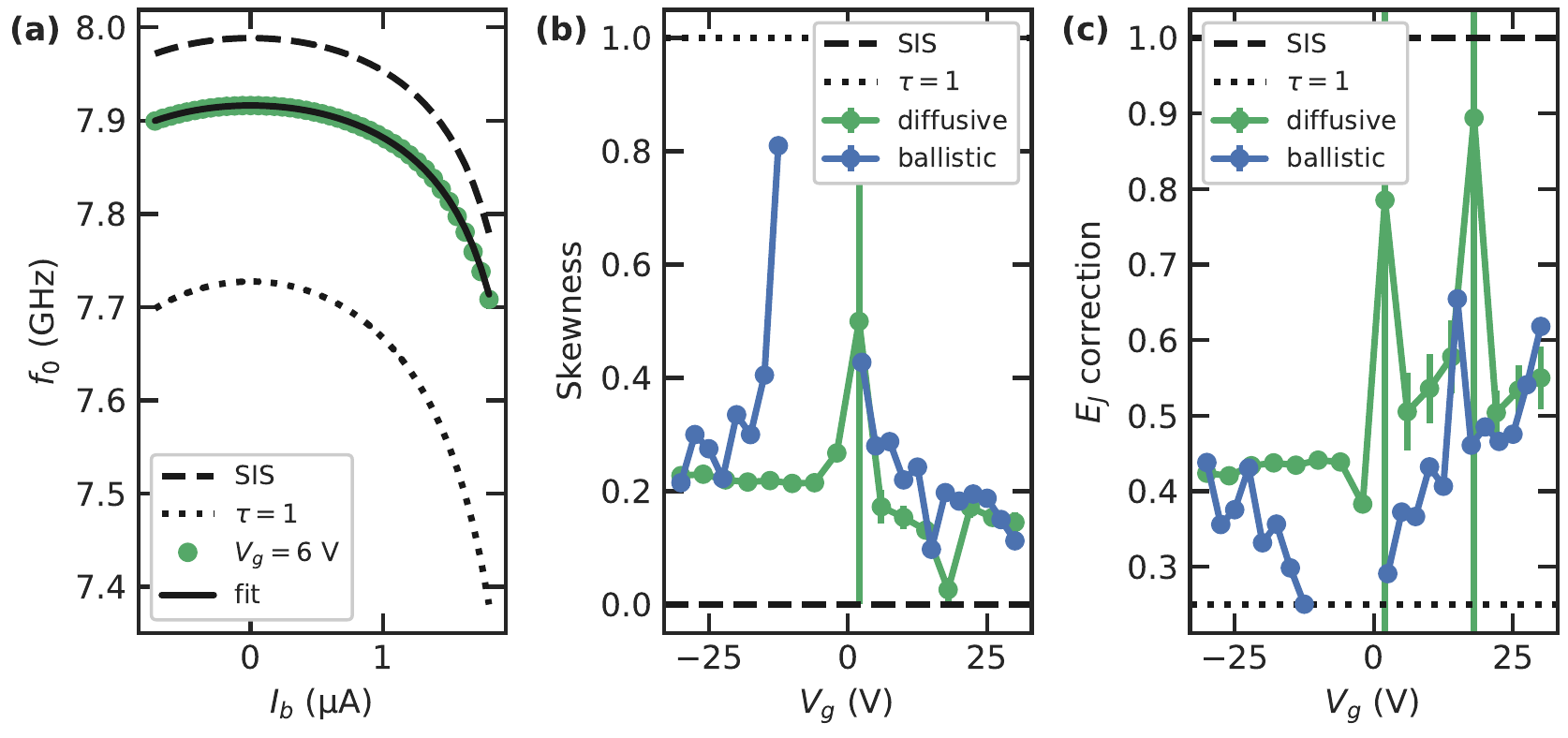}
	\caption{
		\textbf{Observation of the skewness of the current phase relation by measuring the DC current dependence of the linear response of the Josephson inductance.}
		Fitting the bias current dependence \textbf{(a)}, we can extract the junction transparency and corresponding CPR skew \textbf{(b)} for the diffusive (green) and ballistic (blue) gJJ device versus gate voltage.
		A Josephson inductance with underlying SIS-CPR would result in high $f_0$ and small frequency tuning, while maximum skew provides a lower bound on $f_0$.
		\textbf{(d)} Using $\tau$, we calculate the correction factor to $E_\text{J}$ following Eq.~\eqref{CPReq:Ejcorrection} for both devices, indicating significant forward skewing in both samples.
		Dashed lines: Underlying sinusoidal CPR, dotted lines: maximally skewed CPR with $\tau=1$ (see Supplementary Material Fig.~\ref{CPRfig:SMinfluence}).
	}
	\label{CPRfig:figure4}
\end{figure*}

\section{Conclusion}

In summary, we were able to extract evidence of a forward-skewed current phase relation in graphene Josephson junctions by embedding them in superconducting microwave circuits.
Using a combination of drive power and bias current measurements, our results show that scattering of charge carriers, as well as elevated temperature, reduce the CPR skew and with it the circuit anharmonicity via the change in nonlinearity of the JJ itself.

Our circuit architecture is an attractive candidate for analyzing the CPR of exotic JJs, such as ferromagnetic or topological ones~\cite{golubovCurrentphaseRelationJosephson2004a,sochnikovNonsinusoidalCurrentPhaseRelationship2015,stoutimoreSecondHarmonicCurrentPhaseRelation2018,assoulineSpinOrbitInducedPhaseshift2019,muraniMicrowaveSignatureTopological2019}.
Moreover, the influence of high microwave powers on the CPR can be studied straightforwardly, as this only requires repeating the bias current measurements at various powers.
Additionally, the combination of bias current and power dependence should allow to trace out a larger part of the CPR than just around zero phase.

The observed nonlinear damping might unfortunately limit applications of graphene Josephson junctions for cQED.
Devices such as parametric amplifiers need to be operated at high drive powers, which, with nonlinear damping, no longer result in quantum-limited amplification.

%

\section*{Data availability}
All raw and processed data as well as supporting code for measurement libraries, data processing and figure generation is available in Zenodo~\cite{schmidtDataCodeProbing2020}.

\section*{Acknowledgements}
This project has received funding from the European Union Horizon 2020 research and innovation programme under grant agreement No. 785219 -- GrapheneCore2.


\input{main.bbl}


\pagebreak
\clearpage
\widetext

\setcounter{equation}{0}
\setcounter{figure}{0}
\setcounter{table}{0}
\setcounter{page}{1}
\setcounter{section}{0}

\renewcommand{\thepage}{S\arabic{page}}
\renewcommand{\thesection}{S\Roman{section}}
\renewcommand{\thetable}{S\Roman{table}}
\renewcommand{\thefigure}{S\arabic{figure}}
\renewcommand{\theequation}{S\arabic{equation}}
\renewcommand{\bibnumfmt}[1]{[S#1]}
\renewcommand{\citenumfont}[1]{S#1}

\textbf{\centering\large Supplementary Material:\\}
\textbf{\centering\large Probing the current-phase relation of graphene Josephson junctions using microwave measurements\\}

\vspace{1em}

{\centering\noindent Felix E. Schmidt$^{1}$, Mark D. Jenkins$^{1}$, Kenji Watanabe$^{2}$, Takashi Taniguchi$^{2}$ and Gary A. Steele$^{1}$\\}

{\centering\noindent\em $^{1}$Kavli Institute of NanoScience, Delft University of Technology, Lorentzweg 1, 2628 CJ, Delft, The Netherlands.\\
$^{2}$National Institute for Materials Science, 1-1 Namiki, Tsukuba, 305-0044, Japan.\\}


\section{Classification as diffusive or ballistic JJ}\label{sec:SMballistic}

\begin{figure*}[!h]
	\centering
	\includegraphics[width=.7\linewidth]{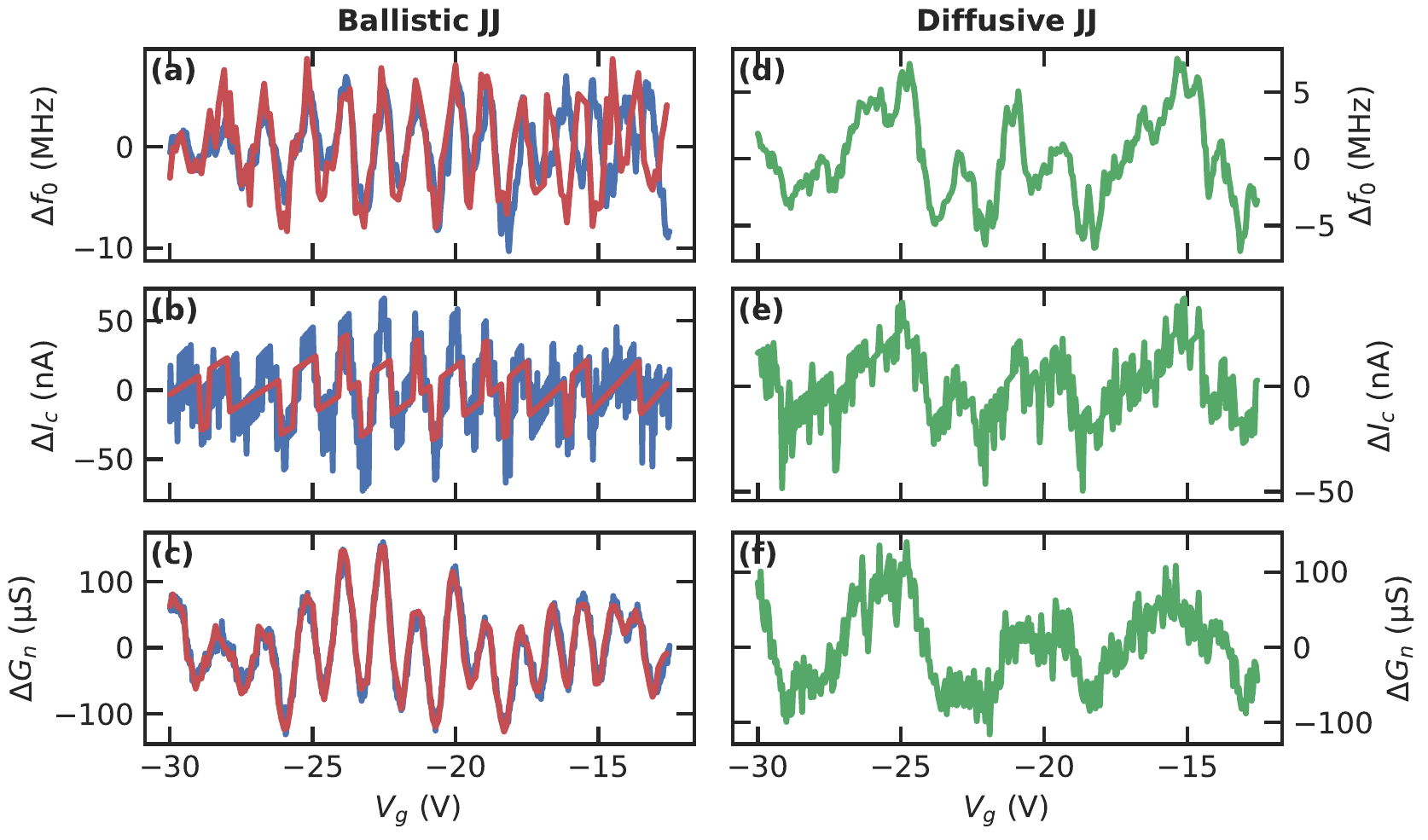}
	\caption{
		\textbf{Fabry-Pérot oscillations in the ballistic device.}
		\textbf{(a-c)} Oscillations in the resonance frequency, DC-switching current and normal state conductance as a function of gate voltage for the ballistic device at base temperature (blue) and \SI{1}{\kelvin} (red).
		\textbf{(d-f)} For the diffusive device, no such features are observed, only a slowly varying background, justifying the classification as \textit{diffusive} device.
	}
	\label{CPRfig:SMFig-ballistic}
\end{figure*}

As stated in the main text, we define the device as ballistic or diffusive in the presence or absence of Fabry-Pérot-like oscillations.
In Fig.~\ref{CPRfig:SMFig-ballistic}, we plot these oscillations after removing a third order background from the data to remove the overall gate-voltage tuning dependence.
Both at base temperature and at \SI{1}{\kelvin}, we observe high-frequency, highly correlated oscillations in all of $f_0$, $I_\text{c}$ and $G_\text{n}=R_\text{n}^{-1}$ for the \textit{ballistic} device, which justifies its classification as such.
The oscillation period allows an estimate of a cavity length of \SI{390}{\nano\meter} for the ABS inside the JJ~\cite{schmidtBallisticGrapheneSuperconducting2018}.
For the same voltage range, however, the \textit{diffusive} device only shows a low-frequency trend originating from the deviation about the removed background, thus lacking the ballistic feature.

\section{Estimation of the fridge attenuation}\label{sec:SMattenuation}

We can estimate the attenuation of our MW input line by using the cryogenic HEMT as a calibrated noise source.
The HEMT noise power is given by
\begin{align}
	P_{\rm HEMT}=10\log\left(\frac{k_\text{B} T_{\rm HEMT}}{\si{\milli\watt}}\right) + 10\log\left(\frac{\Delta f}{\si{\hertz}}\right)\ ,
	\label{CPReq:HEMT}
\end{align}
with the Boltzmann constant $k_\text{B}$, the noise temperature of the HEMT $T_{\rm HEMT}=\SI{2}{\kelvin}$ as specified by the manufacturer and the measurement bandwidth $\Delta f=\SI{100}{\hertz}$.
The resulting noise power is $P_{\rm HEMT}=\SI{-175.59}{dBm}$.
Additionally, we can calculate the average background signal arriving at the VNA by averaging all $S_{11}$ traces in the areas off-resonant to the cavity, which leaves the background unaltered in power.
Doing so, we extract an average signal and standard deviation, which yields the signal-to-noise ratio at the VNA, $\text{SNR}_\text{VNA}=\SI{43.85}{\decibel}$, for a VNA output power of \SI{-20}{dBm}.
Assuming \SI{2}{\decibel} of cable loss between sample and HEMT, we arrive at an attenuation of \SI{111.74}{dB} of our VNA input line,

\section{Extracting $I_\text{c}$ and $f_0$}\label{sec:SMextraction}

The DC switching current (Fig.~\ref{CPRfig:figure1}(b,c)) is taken as the current at which $\partial V/\partial I_\text{b}$ is maximum, where $V$ is the measured voltage drop across the JJ.
Noise or interference on the DC lines could lead to a reduction of the measured $I_\text{c}$ compared to the true value.
To get a more accurate estimation of $I_\text{c}$ together with a good understanding of the noise sources, switching histograms are the preferred measurement method.
The necessary setup was however not available at the time of measurement.

To extract resonance frequency and loss rates from the MW data, we fit the reflection coefficient to the following model (see Ref.~\cite{bosmanBroadbandArchitectureGalvanically2015c} for a derivation):
\begin{align}
	S_{11}(\omega) = -1+\frac{2\kappa_\text{e}}{\kappa+2i\Delta},
\end{align}
where $\kappa=\kappa_\text{e}+\kappa_\text{i}$ denoting the total, external and internal loss rates, respectively, and $\Delta=\omega-\omega_0$ with resonance frequency $\omega_0=2\pi f_0$.
The measured $S_{11}$ is usually distorted by a setup-related microwave background of the following shape:
\begin{align}
	B(\omega) = \left(a+b\omega+c\omega^2\right)e^{i\left(a^\prime+b^\prime\omega\right)},
\end{align}
and with additional rotation by angle $\theta$ in the complex plane, the measured $S_{11}^\prime$ is:
\begin{align}
	S_{11}^\prime(\omega)=B(\omega)\left(e^{i\theta}\left(S_{11}(\omega)+1\right)-1\right)
\end{align}
The origin of the microwave background and phase rotations are impedance mismatches in the wiring originating from various non-ideal circuit elements (e.g. connectors, attenuators, directional couplers, wirebonds).
Standing waves can form in some segments of the wiring which interfere with the measured signal, thus producing an oscillating measurement background.
To remove this background for the gate voltage sweeps (Fig.~\ref{CPRfig:figure1}(d,e)), we pick the measurement trace at the CNP as the one with only background signal, as the MW resonance is extremely broad and effectively not present here.
We then divide the other traces by this trace, resulting in a much cleaner signal.
For measurements based on bias current sweeps, see Fig~\ref{CPRfig:figure4}(a), we take the MW background as the $S_{11}$ trace at $I_\text{b}>I_\text{s}$.
Here, the JJ switched to the normal state and the MW resonance is not present in the measurement.
In order to remove MW background from the power dependence, we mask the regions in which there are resonances for the various powers and gate voltage setpoints, and average the remaining traces.
This way, we obtain a power and frequency map of the MW background, which we use for removing background signal from power traces, such as the one in Fig.~\ref{CPRfig:figure3}(a).

\section{Extracting $f_\text{r}$, $L_\text{r}$ and $L_\text{J}$}\label{sec:SMcalibration}

\begin{figure*}
	\centering
	\includegraphics[width=0.5\linewidth]{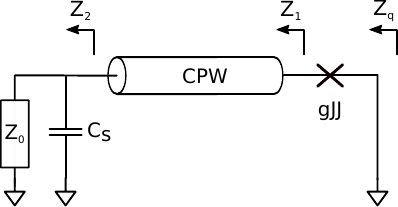}
	\caption{
		\textbf{Derivation of resonance frequency.}
		We define the three impedances $Z_1$, $Z_2$ and $Z_\text{q}$ as seen from the CPW towards the input port, from the gJJ towards the CPW, and as the parallel circuit impedance.
		The gJJ can further be modeled via an RCSJ-model, and an additional gate capacitance (not shown, see text for details).
	}
	\label{CPRfig:rfderivation}
\end{figure*}

We can derive an expression for the circuit resonance frequency depending on the other parameters by using the impedances defined in Fig.~\ref{CPRfig:rfderivation}.
The circuit impedance as seen from the JJ towards the CPW, $Z_1$, the input impedance as seen from the CPW towards the input port, $Z_2$, and the overall parallel circuit impedance $Z_\text{q}$ are:
\begin{align}
	Z_1 &= Z_0 \frac{Z_2+Z_0\tanh\gamma l}{Z_0+Z_2\tanh\gamma l} \\
	Z_2 &= \left(\frac{1}{Z_{C_\text{s}}}+\frac{1}{Z_0}\right)^{-1} = \left(i\omega C_\text{s}+\frac{1}{Z_0}\right)^{-1} \\
	Z_\text{q} &= \left(\frac{1}{Z_\text{J}}+\frac{1}{Z_1}\right)^{-1} = \left(\frac{1}{i\omega L_\text{J}}+\frac{1}{Z_1}\right)^{-1}\ ,
\end{align}
with the CPW length $l$, the complex CPW loss per unit length $\gamma=\alpha+i\beta$, and the transmission line impedance $Z_0$.
Note that the junction impedance $Z_\text{J}$ can be further extended by an RCSJ model and should include additional capacitance for the gate and inductance for the contact electrodes, as described in Ref.~\cite{schmidtBallisticGrapheneSuperconducting2018}.
Assuming negligible losses in the CPW on resonance, $\gamma l\approx i\beta l = i\pi\omega_0/\omega_\text{r}$, i.e. the CPW only acts as a phase shifter.
The resonance condition of the above circuit is for the imaginary part of the admittance $Y=1/Z_\text{q}$ to be zero, which yields
\begin{align}
	0 = \Im \left[ \frac{1}{i\omega_0 L_\text{J}} + \frac{1}{Z_0}\frac{Z_0+iZ_2\tan\left(\pi\omega_0/\omega_\text{r}\right)}{Z_2+iZ_0\tan\left(\pi\omega_0/\omega_\text{r}\right)}\right]
	\label{CPReq:SolAnalytical}
\end{align}
%
%
We can approximate the above by a similar method as the authors of Refs.~\cite{wallquistSelectiveCouplingSuperconducting2006a,wustmannParametricResonanceTunable2013,pogorzalekHystereticFluxResponse2017}:
Assuming a large shunt capacitance at the input, such that $Z_2\approx 0$ and expanding the tangent, we arrive at the expression stated in Eq.~\eqref{CPReq:Pogorzalek}.
This assumption is justified since $C_\text{s}\approx\SI{27}{\pico\farad}$ for our devices, such that both $Z_2\approx \SI{0.2}{\ohm} \ll Z_0=\SI{50}{\ohm}$.
We find that for all values of $L_\text{J}$, including the range in our experiments, the approximation differs by less than \SI{0.2}{\percent} from the analytical solution (see below).

\begin{figure*}
	\centering
	\includegraphics[width=.7\linewidth]{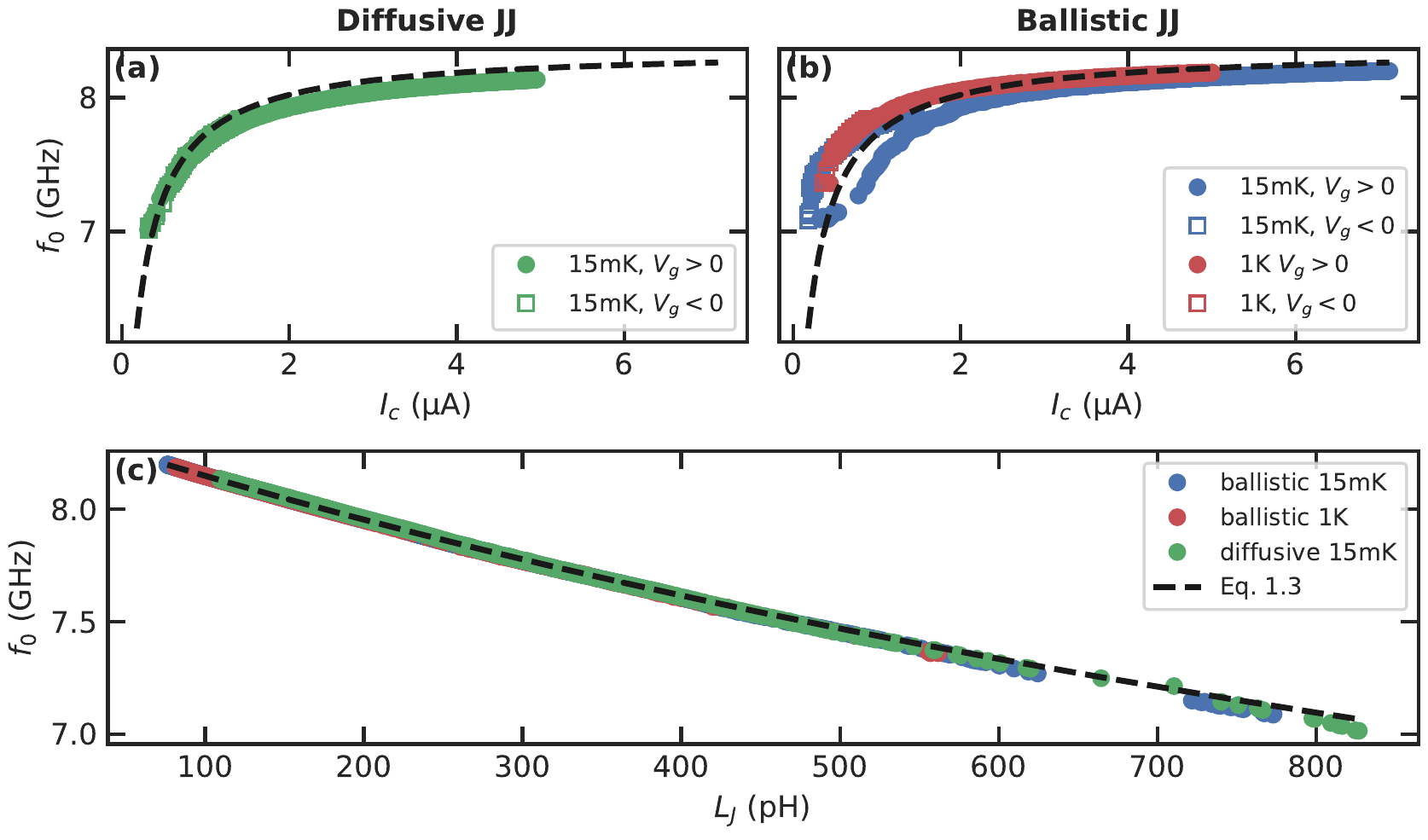}
	\caption{
		\textbf{Resonance frequency vs switching currents for two different gJJ devices.}
		Both the diffusive device at low temperature \textbf{(a)} and the ballistic device at \SI{1}{\kelvin} (\textbf{(b)}, red) show monotonically increasing $f_0$ versus DC-extracted switching currents.
		In contrast, for low temperatures, the ballistic gJJ (\textbf{(b)}, blue) exhibits multi-valued $f_0\left(I_\text{s}\right)$ for gate voltages larger (full circles) and smaller (empty squares) than the charge neutrality point.
		The	multivalued behavior in the ballistic device at low temperature presumably originates from significant differences in junction transparency between n- and p-doping, and only allows for a fit for $V_\text{g}>0$.
		This is not observed at higher temperature or for the diffusive device.
		Dashed lines correspond to Eq.~\eqref{CPReq:Pogorzalek} under assumption of sinusoidal CPR.
		\textbf{(c)} Resonance frequency as a function of observed Josephson inductance, showing good matching to Eq.~\eqref{CPReq:Pogorzalek}.
	}
	\label{CPRfig:SMFigure-f0vsIcvsLj}
\end{figure*}

We extract the circuit parameters from our measurement data in the same fashion as described in the Supplementary Material of Ref.~\cite{schmidtBallisticGrapheneSuperconducting2018}:
In short, we use a reference device with no junction at the end to calibrate $f_\text{r}$ and $L_\text{r}$, a reference device shorted to ground to calibrate the transmission line losses, and finite-element simulations to deduce additional inductances and capacitances of the leads and gate electrode.
This allows us to extract the Josephson inductance directly from the observed resonance frequency, regardless of the underlying CPR.
As shown in Fig.~\ref{CPRfig:SMFigure-f0vsIcvsLj}, while there are significant deviations of Eq.~\eqref{CPReq:Pogorzalek} to the measured $f_0(I_\text{c})$, all measured devices fall on a single curve when plotted as a function of $L_\text{J}$, which verifies this approximation.

\section{Device response to drive power}\label{sec:SMduffing}

\begin{figure*}
	\centering
	\includegraphics[width=.7\linewidth]{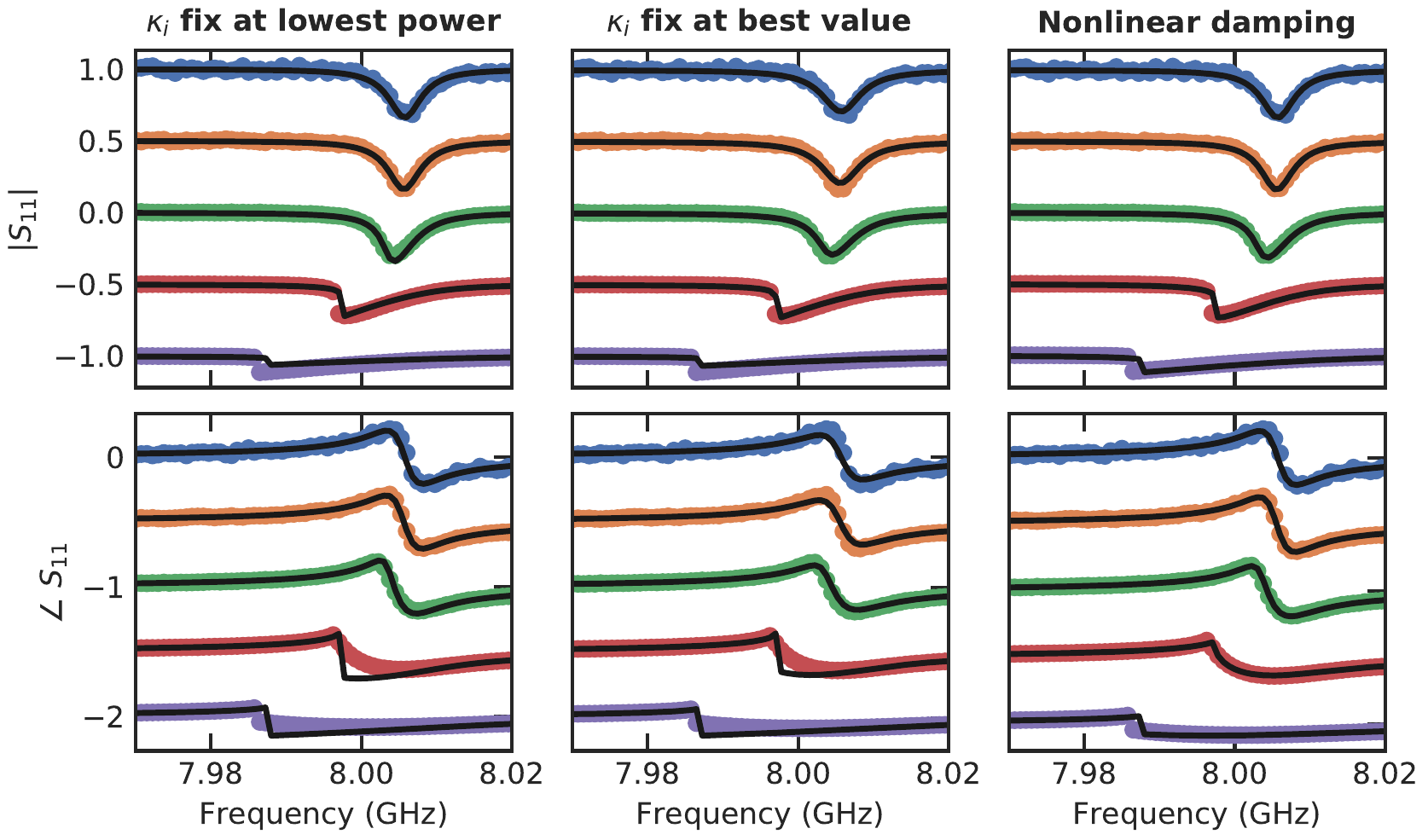}
	\caption{
		\textbf{Anharmonicity fit assuming different cases for $\kappa_\text{i}$.}
		Fixing $\kappa_\text{i}$ to be the value at lowest drive power (first column) results in significantly worse fit than introducing it as constant, but free parameter (second column).
		However, best agreement between data and model is reached when introducing nonlinear damping (third column and Fig.~\ref{CPRfig:figure4}).
		Linecuts and colors correspond to the ones in Fig.~\ref{CPRfig:figure4}.
	}
	\label{CPRfig:SMpower}
\end{figure*}

Following the method described in Ref.~\cite{schmidtCurrentDetectionUsing2020}, the equation of motion of the amplitude field $\alpha(t)$ of a resonator with weak anharmonicity $\beta$ written in the frame rotating with the drive $S_{\rm in}$ is given by Eq.~\eqref{CPReq:Duffing-EOM}, from which the steady-state solution $\partial\alpha_0/\partial t=0$ results in the polynomial function
\begin{align}
	\beta^2 \alpha_0^6 + 2\Delta\beta\alpha_0^4 + \left(\Delta^2+\frac{\kappa^2}{4}\right)\alpha_0^2 - \kappa_\text{e} \abs{S_{\rm in}}^2 = 0\ ,
	\label{CPReq:polynom}
\end{align}
which we can solve and use to calculate the expected reflection coefficient as our model,
\begin{align}
	S_{11}=-1-\frac{\sqrt{\kappa_\text{e}}}{S_{\rm in}}\alpha_0\ .
	\label{CPReq:S11anh}
\end{align}
to fit the measurement data.
We reduce the number of free parameters of this function from five to two by fixing $\omega_0$ and $\kappa_\text{e}$ as the values extracted at lowest drive power and calculating $S_{\rm in}$ from the fridge attenuation, see Supplementary Section Sec.~\ref{sec:SMattenuation}.
The remaining parameters are $\beta$ and $\kappa_\text{i}$, where the internal loss rate can in fact depend on the drive power, $\kappa_\text{i}=\kappa_\text{i}(S_{\rm in})$.
Fixing the loss rate to be constant throughout the fit does not lead to a good fit to the data, as shown in Fig.~\ref{CPRfig:SMpower}.
Our algorithm first fits the measured data to return constant $\beta$ and $\kappa_\text{i}$, and uses these as initial values for a fit to extract the power dependent loss rate.

\begin{figure*}
	\centering
	\includegraphics[width=.7\linewidth]{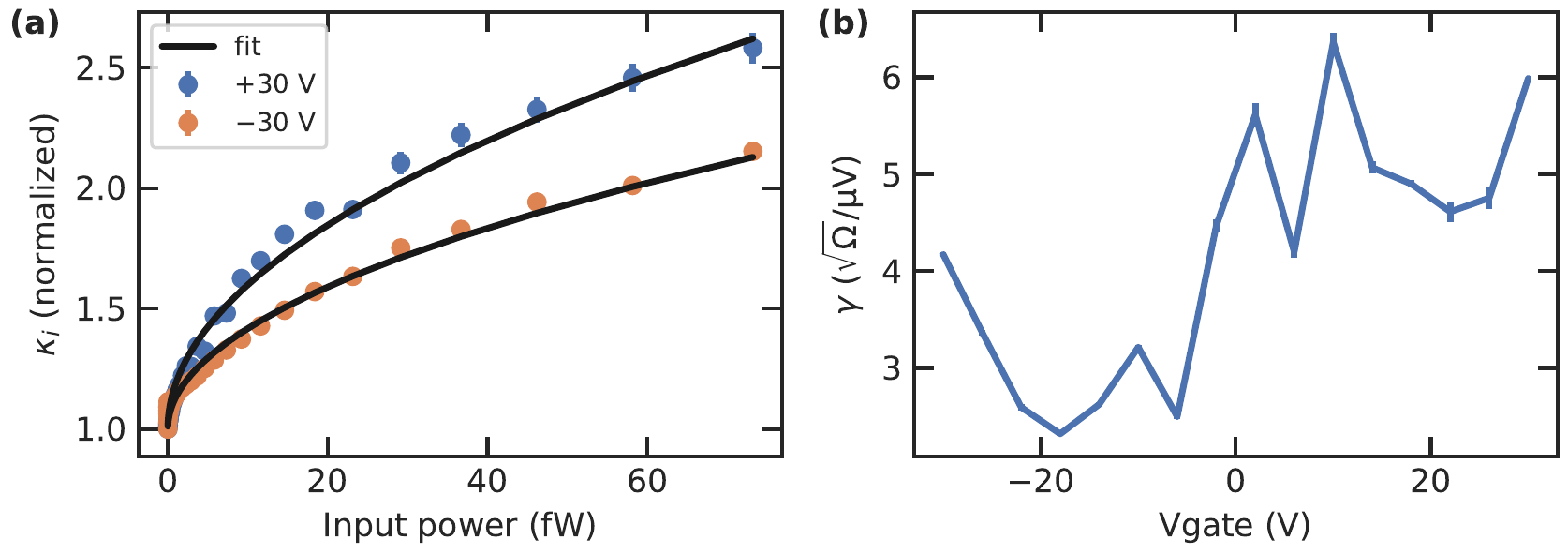}
	\caption{
		\textbf{Nonlinear damping in the gJJ.}
		\textbf{(a)} The internal linewidth of the diffusive device grows with the square root of the input power, regardless of gate voltage.
		\textbf{(b)} The extracted fit parameter $\gamma$ is slightly lower for p-doping compared to n-doping.
		$\gamma$ is related to the subgap losses.
	}
	\label{CPRfig:SMFig-lossrates-power}
\end{figure*}

We can fit the thus extracted change in internal linewidth using a linear growth in drive field $S_{\rm in}$ or square-root dependence on drive power,
\begin{align}
	\kappa_\text{i}=\kappa_\text{i}(0)\left(\gamma\sqrt{P_{\rm in}}+1\right)
	\label{CPReq:kintfitpower}
\end{align}
as shown in Fig.~\ref{CPRfig:SMFig-lossrates-power}(a).
This strongly suggests internal losses originating from sub-gap states populated by the drive field.
Over the range of measured gate voltages, the increase in loss is roughly constant, with slightly larger values for positive compared to negative gate voltages, see Fig.~\ref{CPRfig:SMFig-lossrates-power}(b).

\begin{figure*}
	\centering
	\includegraphics[width=.7\linewidth]{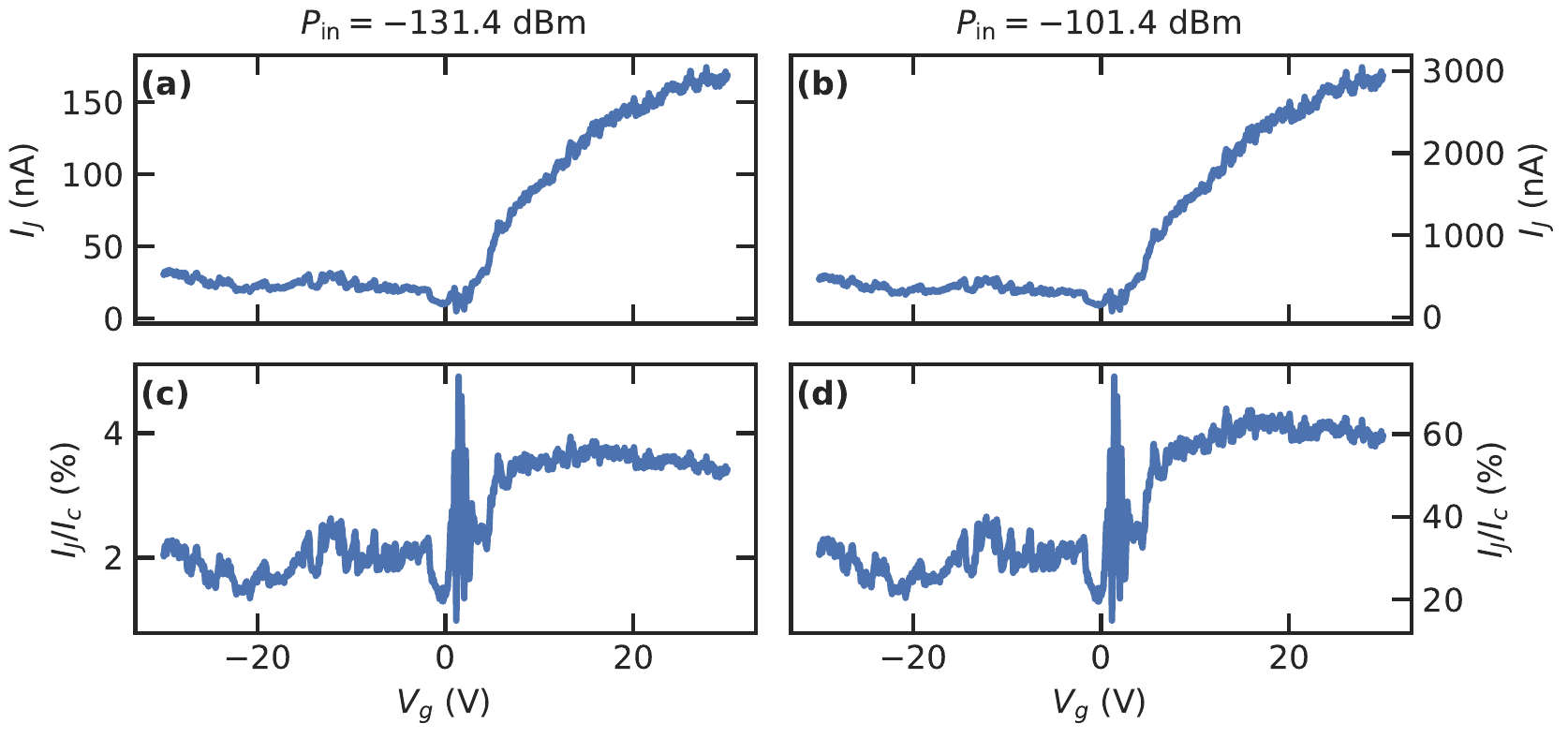}
	\caption{
		\textbf{Current across the diffusive graphene Josephson junction.}
		\textbf{(a,b)} Current across the JJ for varying gate voltage at reference power \textbf{(a)} and maximum drive power \textbf{(b)}, calculated via Eq.~\eqref{CPReq:curratJJ}.
		\textbf{(c,d)} Ratio of current across the JJ to DC-measured switching current for varying gate voltage at reference power \textbf{(c)} and maximum drive power \textbf{(d)}.
		Note the different scales for the left and right column.
	}
	\label{CPRfig:SMFigcurratJJ}
\end{figure*}

Following Ref.~\cite{schmidtCurrentDetectionUsing2020}, we can approximate the current across the junction via the intracavity photon number when driving the device on resonance by combining the input power together with the total and external cavity linewidths:
\begin{align}
	I_0 = \sqrt{\frac{16 P_\text{in} \kappa_\text{e}}{L_\text{r} \kappa^2}}
	\label{CPReq:curratJJ}
\end{align}
In the high-power regime, we estimate the internal linewidths growing according to Eq.~\eqref{CPReq:kintfitpower}, with the coefficient $\gamma$ averaged over all gate voltages.
In Fig.~\ref{CPRfig:SMFigcurratJJ}, we show the estimated currents at the diffusive gJJ for low and high MW powers.
While in the case of low powers (all measurements except for the once in Fig.~\ref{CPRfig:figure3}) the current at the junction is much smaller than $I_\text{c}$, for large drive powers we begin to sample a greater region of the CPR.

\section{Device response to bias current}\label{sec:SMfitbiascurrent}
\subsection{Increasing loss rate}
\begin{figure*}
	\centering
	\includegraphics[width=.7\linewidth]{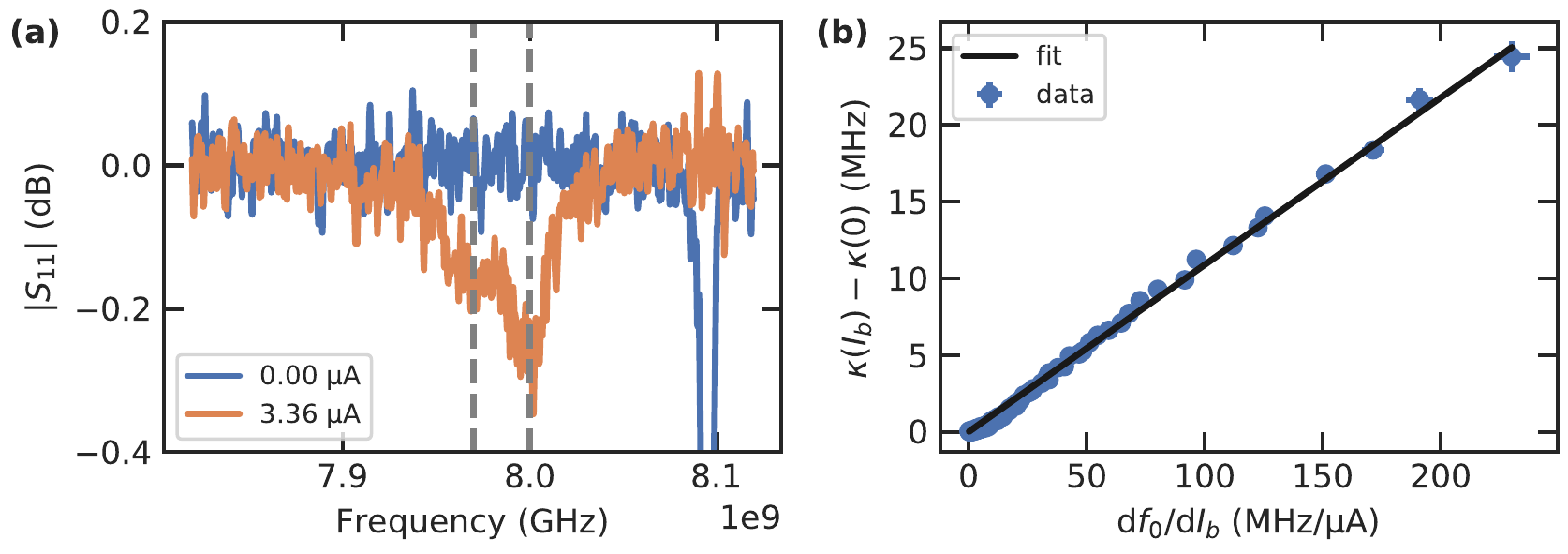}
	\caption{
		\textbf{Internal loss rate for increasing bias current.}
		\textbf{(a)} Compared to the case of zero bias (blue), large bias currents lead to a splitting of the reflected signal in two separate dips (orange).
		The peak spacing and eq.~\eqref{CPReq:currnoise-a}, we can extract a current noise of approximately \SI{270}{\nano\ampere}.
		\textbf{(b)} Difference in total loss rate compared to zero bias current shows a linear increase as a function of responsivity $G_1$, which can be fitted using eq.~\eqref{CPReq:currnoise-b}.
		Increasing loss rate with bias current could originate from  low-frequency noise and/or phase slip events.
	}
	\label{CPRfig:SMFig-lossrates-current}
\end{figure*}

In addition to an increase in $\kappa_\text{i}$ for high drive powers as discussed in the main text, the internal loss rate of our circuit also depends on bias current.
We observe an increasing loss rate for increasing bias current, see Fig~\ref{CPRfig:SMFig-lossrates-current}.
Possible origins of this phenomenon are low-frequency noise on the DC electronics, as this artificially widens the measured cavity resonance if the measurement time is greater than the inverse noise frequency.
Additionally, phase-slip events might occur at larger rates if the Josephson energy potential is tilted, as compared to zero bias current.

The current noise amplitude can be calculated in two ways:
As shown in Fig.~\ref{CPRfig:SMFig-lossrates-current}(a), the reflected signal exhibits a double-peak for bias currents close to $I_\text{c}$, in addition to an increase in linewidth.
This strongly suggests low-frequency current noise, modulating the resonance about the fixed bias current faster than the measurement scan.
From the peak spacing and the measured responsitvity $G_1=\partial f_0/\partial I_\text{b}$, i.e. the change in resonance frequency versus bias current, we can compute the current noise as
\begin{align}
	\Delta I_\text{n} = \frac{\Delta f_0}{\left( \frac{\partial f_0}{\partial I_\text{b}} \right)}
	\label{CPReq:currnoise-a}
\end{align}
From this, we estimate $\Delta I_\text{n}\approx\SI{270}{\nano\ampere}$ due to low-frequency noise for the ballistic device.

Similarly, the increase in total linewidth can be fitted as a linear function of $G_1$, 
\begin{align}
	\kappa_\text{i}(I_\text{b})=\kappa_\text{i}(0)+I_\text{n} \frac{\partial f_0}{\partial I_\text{b}} \ ,
	\label{CPReq:currnoise-b}
\end{align}
resulting in an upper bound for the total corresponding bias current induced losses, see Fig.~\ref{CPRfig:SMFig-lossrates-current}(b).
For the ballistic device, we extract a total corresponding current noise $I_\text{n}\approx\SI{390}{\nano\ampere}$ for the ballistic, and $I_\text{n}\approx\SI{110}{\nano\ampere}$ for the diffusive device.
This leads us to believe that the setup used for the ballistic device was better isolated against current noise than the one for the diffusive device.
Still, some contribution due to processes such as phase slip events is necessary to explain the excess noise obtained from the increase in total linewidth.

\begin{figure*}
	\centering
	\includegraphics[width=.7\linewidth]{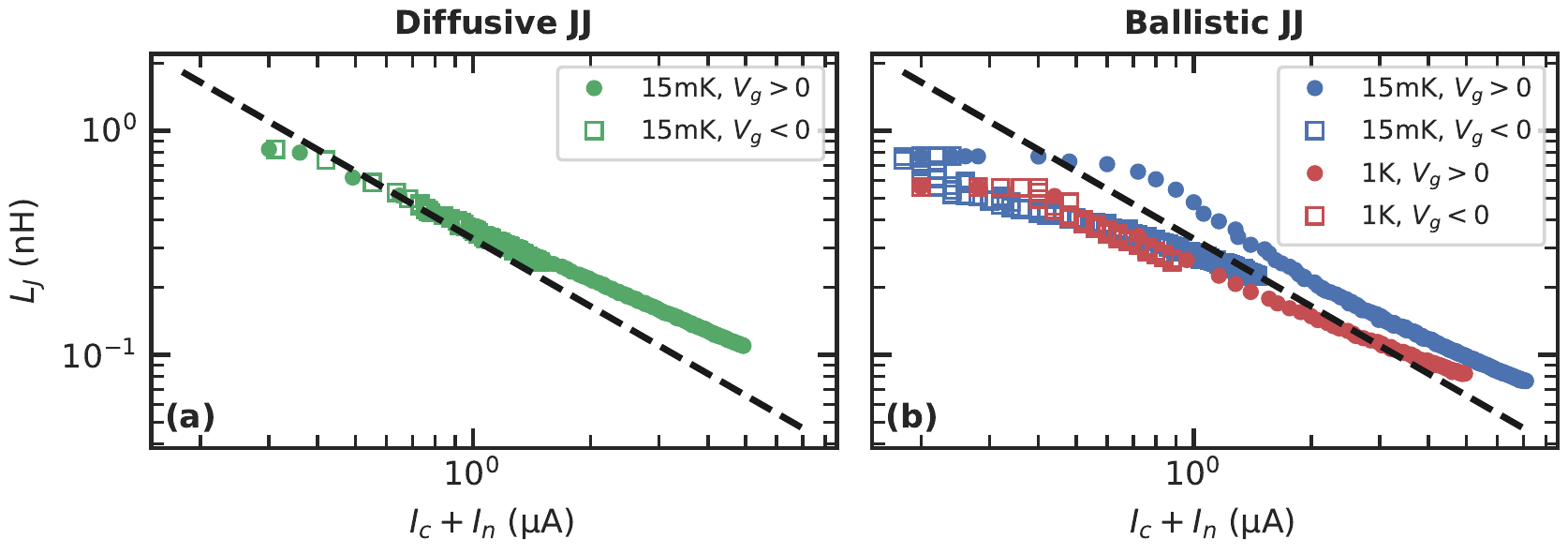}
	\caption{
		\textbf{Josephson inductance and critical currents without added current noise.}
		Without accounting for DC current noise, a significant portion of the measured $L_\text{J}$ drop below the SIS limit of a sinusoidal CPR.
	}
	\label{CPRfig:SMfigure2}
\end{figure*}

Since this current noise also leads to an artificial reduction in the measured $I_\text{c}$, this leads to a rescaling of the current axis of Fig.~\ref{CPRfig:figure2}.
Adding the respective estimates of $I_\text{n}$ to the measured $I_\text{c}$ results in Fig.~\ref{CPRfig:SMfigure2}.
In this case, all measured values of $L_\text{J}$ are larger than the ones extrapolated from $I_\text{c}$ and a sinusoidal CPR, hinting at an overall forward skewed CPR over the full gate voltage range in both devices.
Additional measurements in the form of statistics on the switching current~\cite{kiviojaWeakCouplingJosephson2005a,coskunDistributionSupercurrentSwitching2012,leeProximityCouplingSuperconductorgraphene2018a} could result in more information on this matter, but were not performed at the time.

\subsection{Extracting $\tau$}

\begin{figure*}
	\centering
	\includegraphics[width=.7\linewidth]{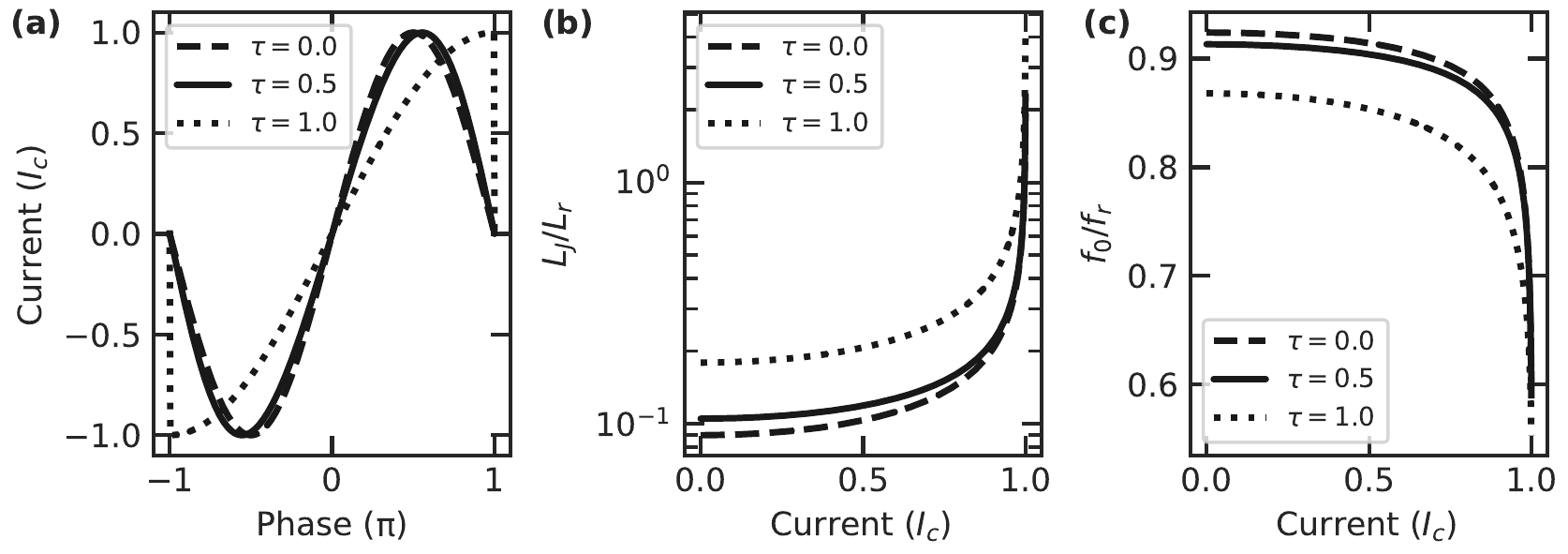}
	\caption{
		\textbf{Predicted influence of the junction transparency on the bias current dependence.}
		\textbf{(a)} CPR for various $\tau=0$ (solid), $\tau=0.5$ (dashed) and $\tau=1.0$	(dash-dotted).
		\textbf{(b-c)} Josephson inductance \textbf{(b)} and resonance frequency \textbf{(c)} normalized to the calibrated values of $L_\text{r}$ and $f_\text{r}$, respectively.
		Each linestyle corresponds to an underlying CPR as calculated in \textbf{(a)}, each with a simulated $I_\text{c}=\SI{1}{\micro\ampere}$.
		Increased forward skewing of the CPR leads to a reduced slope and higher Josephson inductance, which in turn reduces the resonance frequency and increases the tuning.
	}
	\label{CPRfig:SMinfluence}
\end{figure*}

Figure~\ref{CPRfig:SMinfluence} illustrates the effect of a forward skewed CPR on the Josephson inductance and resonance frequency dependence on bias current.
Since a higher skew results in a reduced slope of the CPR, Eq.~\eqref{CPReq:LJgeneral} tells us that $L_\text{J}$ must therefore be increased at zero phase (and current).
Consequently, for the same DC bias microwave circuit with parameters $f_{\lambda/2}$ and $L_\text{r}$, a JJ with larger forward skew and $L_\text{J}$ pushes the initial resonance frequency further downwards than in the case of sinusoidal CPR.

Without any knowledge on the junction transparency $\tau$, fitting data of a CPW cavity with JJ exhibiting a potentially nonsinusoidal CPR can lead to significant deviations from the true circuit parameters.
It is therefore essential to use a fixed set of parameters for $f_{\lambda_2}$ and $L_\text{r}$, as described in Sec.~\ref{sec:SMcalibration}.
To fit the bias current dependence data for extracting $\tau$, we keep these values fixed and only allow $\tau$ and $I_\text{c}$ to vary within reasonable boundaries, i.e. $\tau\in[0,1]$ and $I_\text{c}<\max I_\text{b}$.
Due to the significant current noise, the cavity resonance gets very broad and begins to resemble a double-dip feature, which makes extraction of reliable values for small gate voltages increasingly difficult.
For this reason, we chose to omit gate voltages below the CNP from further analysis.

\input{SM.bbl}
\end{document}

%% file: main.bbl
%